
\magnification=1200
\vsize=7.5in
\hsize=5.6in
\tolerance 500

\baselineskip 13pt plus 1pt minus 1pt

\def\footnoterule{\kern-3pt \hrule width \hsize \kern2.6pt}
\pageno=0
\footline={\ifnum\pageno>0 \hss --\folio-- \hss \else\fi}

\centerline{
{\bf CHIRAL-SYMMETRY AND STRANGE FOUR-QUARK MATRIX ELEMENTS}\footnote{*}
{This work is supported in part by funds
provided by the U.S. Department of Energy
(D.O.E.) under contract \#DE-AC02-76ER03069.}}
\vskip 24pt
\centerline{Xiangdong Ji}
\vskip 12pt
\centerline{\it Center for Theoretical Physics}
\centerline{\it Laboratory for Nuclear Science}
\centerline{\it  and Department of Physics}
\centerline{\it Massachusetts Institute of Technology}
\centerline{\it Cambridge, Massachusetts 02139}
\vfill
\centerline{\bf ABSTRACT}
\bigskip
\midinsert
\baselineskip 24pt plus 1pt minus 1pt
\narrower
\noindent
We consider the matrix elements of the left-handed
flavor-conserving four-quark operators in
the nucleon and pion states.
Using chiral symmetry, we derive relationships
among these matrix elements.
We argue that the $\Delta I = 1/2$ rule of
hyperon and kaon non-leptonic weak decay implies possible
large strange-quark content in the
nucleon and pion.
\endinsert
\vfill
\centerline{Submitted to: {\it Physics Letters B}}
\vfill
\line{CTP\#2188 \hfil hep-ph/9307234 \hfil February 1993}
\eject

The nucleon's strange content has received
considerable attention since the EMC measurement of
the proton's spin structure function $g_1(x)$.$^1$
Many more deep-inelastic and (quasi)-elastic
experiments are now in running or have been
proposed to make further measurement of
strange-quark matrix elements.$^2$ Study
of the nucleon's non-valence degrees of
freedom helps us to understand better
the role played by the sea of Quantum
Chromodynamics (QCD). It also provides
clues for more realistic model-building
and for finding better approximations
to solve QCD.

In this letter, we discuss the strange four-quark
matrix elements in the nucleon and pion states.
Our discussion is motivated by a recent paper
by Kaplan,$^3$ who showed that the well-known
$\Delta I=1/2$ rule of hyperon non-leptonic weak decay
implies large strange four-quark matrix element,
$\langle P|\bar u_L\gamma_\mu s_L\bar
s_L\gamma^\mu d_L|N\rangle$. Our goal here
is to elaborate on his result
from the relationships among four-quark
matrix elements implied by chiral symmetry, and
to show that the vanishing strange-quark matrix elements
are incompatible with the $\Delta I =1/2$ rule.
Then we extend our discussion
to the pion and argue that accommodation of
both the $\Delta I =1/2$ rule and vanishing
strange-quark matrix elements requires a drastically
different valence quark model from what we have constructed.
If there is a trail of pion large strange content
here, this may be the first that the pion has an
intricate flavor structure.

Let us consider the following tensor of
four-quark operators,
$$      T^{ik}_{jl} = {1\over 4}~ \bar q_i\gamma^\mu(1-\gamma_5)
        q_j \bar q_k\gamma_\mu(1-\gamma_5)q_l,  \eqno(1) $$
which will be denoted by $(\bar q_iq_j)
(\bar q_kq_l)$, or simply $\bar q_iq_j\bar q_kq_l$. The indices
$i, j, k, l$ run through light-quark flavors $u$,
$d$, and $s$. Color indices, unless specified explicitly,
are coupled to singlet in the quark pairs ($i$, $j$) and ($k$, $l$).
The tensor is symmetric under simultaneous exchange of
$i$ and $k$, and $j$ and $l$, and thus symmetrizing
$i$ and $k$ and anti-symmetrizing $j$ and $l$, or vice versa,
yield a null result. We have therefore a total of
$3^4 - 3\times 6\times 2 = 45$ independent tensor
components. The
symmetrized tensor $T^{(ik)}_{(jl)}$ has $6\times 6 = 36$
components and the anti-symmetrized tensor $T^{[ik]}_{[jl]}$
has $3 \times \bar 3 = 9$ components. The former
contains 27, 8, and 1 representations of $SU(3)_L$
and the latter 8 and 1 representations.

The construction of the operators belonging to different
$SU(3)_L$ representations is standard. Here we present the
result in order to specify the normalization
we adopt. For 27, we subtract away the trace of the
symmetric tensor,
$$  \eqalign{      \bar T^{(ik)}_{(jl)} = &
            T^{(ik)}_{(jl)} - {1\over 5}\Big[
            T^{(mk)}_{(ml)}\delta^i_j + T^{(mi)}_{(mj)}\delta^k_l +
            T^{(mi)}_{(ml)}\delta^k_j + T^{(mk)}_{(mj)}\delta^i_l\Big]
           \cr & + {1\over 20} T^{(mn)}_{(mn)}\Big[\delta^i_j\delta^k_l
               + \delta^i_l\delta^k_j\Big]. } \eqno(2)    $$
For the symmetric and anti-symmetric octets we define,
$$         O_{S,A}^a = (\bar q_\alpha T^a q_\alpha)
                    (\bar q_\beta q_\beta) \pm
                    (\bar q_\alpha T^a q_\beta)
                    (\bar q_\beta q_\alpha), \eqno(3) $$
where the Gell-Mann matrices $T^a$ are normalized according to
${\rm Tr}T^aT^b = 1/2\delta^{ab}$. The repeated
$\alpha$ and $\beta$ indices denote summation
over color, and the flavor indices are coupled
in each bracket implicitly. For symmetric and anti-symmetric
singlets we define,
$$           S_{S,A} = (\bar q_\alpha q_\alpha)
                    (\bar q_\beta q_\beta) \pm
                    (\bar q_\alpha q_\beta)
                    (\bar q_\beta q_\alpha). \eqno(4) $$

Any tensor component in eq. (1) can be decomposed into
a sum of operators belong to these
five different representations. For diagonal hadron
matrix elements, we are interested in operators with no
net flavor change. There are nine of them:
$\bar uu\bar uu$, $\bar dd\bar dd$, $\bar ss\bar ss$,
$\bar uu\bar dd$, $\bar uu\bar ss$, $\bar dd\bar ss$,
$\bar ud\bar du$, $\bar us\bar su$, and $\bar ds\bar sd$.
Their decomposition into different representations
is straightforward,
$$ \eqalign{  \bar uu\bar uu & = \bar T^{uu}_{uu}  + {2\over 5}
           O_S^3  + {2\over 5\sqrt{3}}O_S^8 + {1\over 12}S_S, \cr
    \bar dd\bar dd & = \bar T^{dd}_{dd}  - {2\over 5}
           O_S^3  + {2\over 5\sqrt{3}}O_S^8 + {1\over 12}S_S, \cr
   \bar ss\bar ss&  = \bar T^{ss}_{ss}  -{4\over 5\sqrt{3}}O_S^8
       + {1\over 12}S_S,}  $$
$$ \eqalign{ {1\over 2}(\bar uu\bar dd + \bar ud\bar du)
     &  = \bar T^{ud}_{ud} + {1\over 5\sqrt{3}}O_S^8 + {1\over 24} S_S,  \cr
   {1\over 2}(\bar dd\bar ss + \bar ds\bar sd)
     &  = \bar T^{ds}_{ds} -{1\over 10}O_S^3 - {1\over 10\sqrt{3}}O_S^8
           + {1\over 24} S_S,  \cr
  {1\over 2}(\bar ss\bar uu + \bar su\bar us)
     &  = \bar T^{su}_{su} + {1\over 10}O_S^3 - {1\over 10\sqrt{3}}O_S^8
           + {1\over 24} S_S, }   $$
$$ \eqalign{  {1\over 2}(\bar uu\bar dd - \bar ud\bar du)
      & = {1\over \sqrt{3}}O_A^8 + {1\over 12} S_A, \cr
   {1\over 2}(\bar dd\bar ss - \bar ds\bar sd)
      & = -{1\over 12 }O_A^3 - {1\over 2\sqrt{3}}O_A^8
           + {1\over 12} S_A,    \cr
    {1\over 2}(\bar ss\bar uu - \bar su\bar us)
      & = {1\over 12}O_S^3 - {1\over 2\sqrt{3}}O_A^8
           + {1\over 12} S_A. }\eqno(5) $$
Likewise, we can decompose the $\Delta s= 1$
non-leptonic weak decay hamiltonian density,
$$    {\cal H} = {G_F\over \sqrt{2}} s_1
              c_1 4(\bar d u)(\bar u s).    \eqno(6) $$
However, this hamiltonian density is defined
at the scale of $M_W$, the mass of the $W$ boson.
Since we are interested in the non-perturbative
part of the four-quark matrix elements, we have to
run down the scale using a renormalization
group equation. The calculation is standard
and a recent reference shows,$^4$,
$$            { \cal H}(\mu = 1 {\rm GeV}) =  {G_F\over \sqrt{2}} s_1
              c_1 \Big[ 2.8 \bar T^{du}_{us}
           + 0.28 O_S^{6+i7} - 3.64O_A^{6+i7}
           + 0.003Q_5 - 0.01Q_6\Big].  \eqno(7) $$
where
$$\eqalign{ Q_5 & = \bar d_\alpha \gamma^{\mu}(1-\gamma_5)
           s_\alpha \bar q_\beta \gamma_\mu(1+\gamma_5) q_\beta,  \cr
       Q_6 & = \bar d_\alpha \gamma^{\mu}(1-\gamma_5)
           s_\beta \bar q_\beta \gamma_\mu(1+\gamma_5) q_\alpha, }
      \eqno(8) $$
are generated from the penguin diagram and are interesting
due to their distinct chiral structure.

We now consider the matrix elements of
those flavor-conserving operators in the nucleon
and pion states. [Since the operators are scale-dependent,
we assume to work at the scale of 1 GeV.] To do that, we first
map these quark operators onto operators containing baryon and
meson fields with the same flavor symmetry. We use the non-linear
representations for mesons and baryons employed in ref. 5: the
Goldstone bosons octet is represented by a $3\times 3$ matrix,
$$             \Sigma = \exp(2i\pi/f_\pi), \eqno(9) $$
where
$$          \pi = {1\over \sqrt{2}}
          \pmatrix{{1\over \sqrt{2}}\pi^0 + {1 \over \sqrt{6}}\eta
          & \pi^+ & K^+ \cr
          \pi^- & -{1\over \sqrt{2}}\pi^0 + {1 \over \sqrt{6}}\eta & K^0 \cr
           K^- & K^0 & -{2\over \sqrt{6}}\eta  }, \eqno(10) $$
and the baryon octet is represented by
$$    B = \pmatrix{{1\over \sqrt{2}}\Sigma^0 + {1 \over \sqrt{6}}\Lambda
      & \Sigma^+ & P \cr
      \Sigma^- & -{1\over \sqrt{2}}\Sigma^0 + {1 \over
      \sqrt{6}}\Lambda & N \cr \Xi^- & \Xi^0 & -{2\over \sqrt{6}}\Lambda  } .
\eqno(11) $$
Under chiral transformation, we have
$$ \eqalign{         \Sigma &  \rightarrow L\Sigma R^\dagger, \cr
           \xi = \sqrt {\Sigma} & \rightarrow L\xi U^\dagger = U\xi R^\dagger,
             \cr B & \rightarrow   UBU^\dagger.  }        \eqno(12) $$
In constructing effective operators, we
keep only leading terms in chiral perturbation expansion.
For 27, we have
$$   \bar T^{(ik)}_{(jl)} \rightarrow a\bar B^{(ik)}_{(jl)}, \eqno(13) $$
where the tensor $B$ is,
$$         B^{(ik)}_{(jl)} = (\xi \bar B\xi^\dagger)^j_i
                    (\xi B\xi^\dagger)^l_k,   \eqno(14)  $$
and for 8's,
$$      O_{S,A}^a \rightarrow F_{S,A} {\rm Tr} \bar B[\xi^\dagger T^a\xi,B]
                   + D_{S,A} {\rm Tr} \bar B[\xi^\dagger T^a\xi, B]_+,
\eqno(15)  $$
and finally for 1's,
$$      S_{S,A} \rightarrow s_{s,a} {\rm Tr} \bar B B.  \eqno(16)  $$
The seven parameters, $a$, $F_{S,A}$, $D_{S,A}$, and $s_{s,a}$
determine all matrix elements of the four-quark operators
between the baryon octet states plus an
arbitrary number of Goldstone bosons.

Using the above mapping, we calculate the matrix elements of the
flavor-conserving operators between the proton states,
$$ \eqalign{   \langle P |\bar uu\bar uu |P\rangle
      &  = -{3\over 20} a + {1\over 5}(F_S + D_S)
         + {1\over 15}(3F_S -D_S) + {s_s \over 12},  \cr
     \langle P |\bar dd\bar dd |P\rangle
     &  = {1\over 20} a - {1\over 5}(F_S + D_S)
         + {1\over 15}(3F_S -D_S) + {s_s \over 12},  \cr
    \langle P |\bar ss\bar ss |P\rangle
      &  =  - {3\over 20}a
         - {2\over 15}(3F_S -D_S) + {s_s \over 12}, } $$
$$ \eqalign{  \langle P |{1\over 2}(\bar uu\bar dd + \bar ud\bar du) |P\rangle
      & = -{1\over 40} a
         + {1\over 30}(3F_S -D_S) + {s_s \over 24},  \cr
   \langle P |{1\over 2}(\bar dd\bar ss + \bar ds\bar sd) |P\rangle
      & = -{1\over 40} a - {1\over 20}(F_S + D_S)
         - {1\over 60}(3F_S -D_S) + {s_s \over 24},  \cr
  \langle P |{1\over 2}(\bar ss\bar uu + \bar su\bar us) |P\rangle
      & = {7\over 40} a + {1\over 20}(F_S + D_S)
         - {1\over 60}(3F_S -D_S) + {s_s \over 24},} $$
$$ \eqalign{  \langle P |{1\over 2}(\bar uu\bar dd - \bar ud\bar du) |P\rangle
     &  =  {1\over 6}(3F_A -D_A) + {s_a \over 12},  \cr
 \langle P |{1\over 2}(\bar dd\bar ss - \bar ds\bar sd) |P\rangle
     &  = - {1\over 4}(F_A + D_A)
         - {1\over 12}(3F_A -D_A) + {s_a \over 12},  \cr
 \langle P |{1\over 2}(\bar ss\bar uu - \bar su\bar us) |P\rangle
     &  = {1\over 4}(F_A + D_A)
         - {1\over 12}(3F_A -D_A) + {s_a \over 12}.} \eqno(17) $$
The coefficients in front of the invariant parameters
are related to the SU(3) Clebsch-Gordon coefficients.

If we assume that five strange quark matrix elements in eq.
(17) vanish, we immediately
derive five relations among seven invariant parameters,
$$ \eqalign{ F_A & = -D_A = {1\over 4}s_a,  \cr
              a& = -{1\over 3} s_s, \cr
             F_S & = {5\over 12} s_s, \cr
             D_S & = {1\over 4} s_s.} \eqno(18) $$
Here we have taken $s_a$ and $s_s$ as independent.
The rest four non-strange matrix elements can be
expressed in terms of these two parameters,
$$ \eqalign{   \langle P|\bar u u \bar uu |P \rangle & = {1\over 3} s_s, \cr
        \langle P|\bar dd\bar dd|P \rangle & = 0,   \cr
       \langle P|{1\over 2}(\bar u u \bar dd
           - \bar ud\bar du) |P \rangle &  = {1\over 4}s_a,   \cr
   \langle P|{1\over 2}(\bar u u \bar dd
           + \bar ud\bar du) |P \rangle & = {1\over 12}s_s.} \eqno(19)   $$
The vanishing of the four-$d$-quark matrix element is a little
surprising, but it can be simply interpreted in
valence quark models in which there is only one $d$ quark and
thus the two-body matrix element must vanish. However, our result
is independent of valence quark models and is directly
linked to the SU(3) symmetry and
vanishing strange-quark matrix elements.
The relation between the four-$u$-matrix element and
the $ud$-symmetric matrix element (the last one in eq. (19))
is simply a consequence of isospin symmetry.

We argue, however, that the pattern of four-quark
matrix elements shown in eqs. (18) and (19)
is inconsistent with the data on $\Delta s=1$ hyperon non-leptonic
decay. To show this, we map the hamiltonian in eq. (7)
onto an operator with meson and baryon fields
and use it to calculate the hyperon
decay rates. Here we neglect the contributions of the
penguin operators because of their small coefficients
(a more careful analysis was made in ref. 3).
{}From fitting to experimental data,$^6$ we have
$$ \eqalign{    0.07F_S - 0.91 F_A & =  1.4 \eta
      {m_\pi^2 f_\pi\over 2s_1c_1},  \cr
    0.07D_S -0.91 D_A & = -0.58 \eta {m_\pi^2 f_\pi\over 2s_1c_1}, }
\eqno(20) $$
and $a$ is negligible. Here $\eta$ is a phase factor,
and time-reversal symmetry restricts $\eta$ to $\pm 1$.
{}From eq. (18), the size of $a$ restricts $F_S$ and $D_S$
to be small, i.e., the symmetric octet is also
strongly suppressed if
all the strange quark matrix elements vanish. Then
from eq. (20), we find the ratio
between $F_A$ and $D_A$ is $-2.4$, which contradicts with
$F_A/D_A = -1$, which is implied by eq. (18).

Thus some strange matrix elements must be large.
{}From eq. (17) and the fact that $a$ is small, we have,
$$     {1\over 5}(F_S + D_S) = \langle P|
          (\bar uu-\bar dd) \bar ss+
          (\bar us\bar su - \bar ds\bar du)|P\rangle,  $$
$$     (F_A + D_A) = \langle P|
          (\bar uu-\bar dd) \bar ss -
          (\bar us\bar su - \bar ds\bar du)|P\rangle.  \eqno(21) $$
Combining these with eq. (20), we have,
$$      \langle P|-0.56(\bar uu - \bar dd)\bar ss
            + 1.26(\bar us\bar su-\bar ds\bar sd) |P\rangle
       = 0.82\eta{m_\pi^2 f_\pi\over 2s_1c_1}  ,  \eqno(22) $$
which is the matrix element obtained by Kaplan,$^3$
except for an isospin rotation.

In the rest of the letter, we show that
the above line of discussion can be extended to
the pion's matrix elements, although
some arguments for the conclusion are
less tight. In the Goldstone boson sector, we have
the following mapping for the four quark matrix elements:
$$       \bar T^{(ik)}_{(jl)}
        \rightarrow  a{f_\pi^2\over m_\pi^2} \bar
           P^{(ik)}_{(jl)},  \eqno(23) $$
where the tensor $P^{(ik)}_{(jl)}$ is defined as,
$$        P^{(ik)}_{(jl)} = (\Sigma \partial^\mu \Sigma^\dagger)^j_i
              (\Sigma \partial_\mu \Sigma^\dagger)^l_k, \eqno(24) $$
and
$$       O_{S,A}^a \rightarrow b_{s,a} {f_\pi^2\over m_\pi^2}{\rm Tr}
              [T^a \partial_\mu \Sigma \partial^\mu \Sigma^\dagger], $$
$$       S_{S,A} \rightarrow s_{s,a}{f_\pi^2\over m_\pi^2}
           {\rm Tr} [\partial_\mu \Sigma \partial^\mu \Sigma^\dagger],
     \eqno(25)  $$
where we introduce five invariant parameters, $a$, $b_{s,a}$,
and $s_{s,a}$. Using these, we obtain the following
four-quark matrix elements of the pion,
$$ \eqalign{     \langle \pi^0|\bar uu\bar uu|\pi^0 \rangle
      & = -{7\over 40}a + {1\over 30}b_s + {1\over 12} s_s,  \cr
       \langle \pi^0|\bar ss\bar ss|\pi^0 \rangle
      & = -{1\over 40}a - {1\over 15}b_s + {1\over 12} s_s,  \cr
      \langle \pi^0|{1\over 2}[\bar uu\bar dd
         + \bar ud \bar du] |\pi^0 \rangle
      & = {13\over 80}a + {1\over 60}b_s + {1\over 24} s_s,  \cr
   \langle \pi^0|{1\over 2}[\bar uu\bar ss
         + \bar us \bar su] |\pi^0 \rangle
      & = {1\over 80}a - {1\over 120}b_s + {1\over 24} s_s,  \cr
      \langle \pi^0|{1\over 2}[\bar uu\bar dd
        - \bar ud \bar du] |\pi^0 \rangle
      & = {1\over 12}b_a + {1\over 12} s_a,   \cr
      \langle \pi^0|{1\over 2}[\bar uu\bar ss
         - \bar us \bar su] |\pi^0 \rangle
      & =  - {1\over 24}b_a + {1\over 12} s_a.  } \eqno(26) $$
The matrix elements of three other operators, $\bar dd\bar dd$,
$\bar dd\bar ss$, and $\bar ds \bar sd$, are redundant because
of isospin symmetry.

If we assume all strange-quark matrix elements vanish, we
derive three relations among five invariant parameters,
$$ \eqalign{           b_a & = 2s_a, \cr
                  b_s & = -a = 2s_s. } \eqno(27)  $$
Three non-strange matrix elements are,
$$ \eqalign{       \langle \pi^0| \bar uu\bar uu|\pi^0\rangle & = {1\over 2}
         s_s, \cr
          \langle \pi^0|{1\over 2}
           [\bar uu\bar dd + \bar ud \bar du ]|\pi^0\rangle & = -{1\over 4}
          s_s, \cr
         \langle \pi^0|{1\over 2}
           [\bar uu\bar dd - \bar ud \bar du ]|\pi^0\rangle & = {1\over 4}
        s_a.}
     \eqno(28) $$
The first two matrix elements are again related by isospin rotation.

We now argue that eqs. (27) and (28)
and the $\Delta I =1/2$ rule imply either
a non-conventional valence quark model or a large strange
content in the pion.
{}From the neutral and charged $K$ non-leptonic
decay, we can extract the strengths
of the 27 and 8,$^5$
$$ \eqalign{        a & = 1.4 \times 10^{-5}\eta {\rm GeV}^4,  \cr
            0.07b_s-0.91b_a & = 2.14 \times 10^{-4}\eta' {\rm GeV}^4. }
         \eqno(29) $$
To be more precise, the second equation above shall include
the penguin contributions which are potentially important in the meson
sector due to the special chiral structure of $Q_5$ and $Q_6$.$^7$
Since their precise size at 1 GeV is uncertain, we neglect it temporarily.
Coupling eq. (29) with eq. (27) we deduce,
$$ \eqalign{   s_s&  = -0.7 \times 10^{-5}\eta {\rm GeV}^4,   \cr
               s_a&  = -1.2 \times 10^{-4}\eta' {\rm GeV}^4,} \eqno(30)$$
and so $|s_a/s_s|\sim 17$. Or, in terms of a ratio of
non-strange matrix elements,
$$        \chi =  { \langle \pi|\bar ud\bar du |\pi \rangle \over
             \langle \pi|\bar uu\bar dd |\pi \rangle} \approx -1. \eqno(31) $$

One way to estimate the penguin contributions is to
consider a chiral theory in the large $N_c$
(the number of color) limit. It was shown in ref. 8 that
the corresponding effective operator for $Q_6$ is
$$       -({f_K\over f_\pi} -1) {4m_K^2m_\pi^2f_\pi^2 \over m_s^2}
{f_\pi^2\over m_\pi^2}{\rm Tr}
             [T^a \partial_\mu \Sigma \partial^\mu \Sigma^\dagger].
      \eqno(32) $$
This contributes a term of $5.3\times 10^{-6}{\rm GeV}^4$ to the
left hand side of the second equation in (29).
This is clearly too small to bring $b_s$ and $b_a$ to
a same size. It is known that the large $N_c$ method is
not always reliable$^8$, however, for the penguin to
explain the $\Delta I =1/2$ rule, the realistic penguin
matrix element must be 40 times the large $N_c$
result. A recent lattice calculation
shows that the large $N_c$ estimate is correct in order
of magnitude at least for the $K\rightarrow \pi$ matrix element.$^9$
In light of this, we take eq. (31) as qualitatively
true under the assumption about the strange matrix elements.

This result, however, contradicts various valence-quark models
for the pion. For instance, in the non-relativistic
quark model, we have,
$$\chi = 1/3. \eqno(31) $$
The same result can also be obtained in the vacuum insertion
approximation, in which the matrix element of
four-quark operators are calculated
by inserting a physical vacuum in the middle.$^7$
In the MIT bag model, we find,
$$        \chi = {1\over 3} {\int (j_0^4 +j_1^4 - {2\over 3}j_0^2j_1^2)
                  \over \int (j_0^2 + j_1^2)^2},  \eqno(32)  $$
where $j_0$ and $j_1$ are upper and lower components of the
bag wave function. Clearly, for any $j_0$ and $j_1$,
$\chi$ is larger than zero, but smaller than 1/3. Thus
either different four-quark operators in these models acquire
different renormalization constants, or the strange degrees of
freedom must be added explicitly, or both.

If the above discrepancy implies a large strange
content of the pion, it is the first such evidence.
To support this claim, let us consider the
matrix elements of a few familar strange quark operators
in the pion. The matrix elements of type $\langle
0|\bar s\Gamma s|\pi^0\rangle $ clearly vanish
due to isospin symmetry. The matrix elements
of type $\langle \pi^0|\bar s\Gamma s|\pi^0\rangle $
vanish exactly except for $\Gamma = 1$.
However, the chiral symmetry predicts
$\langle \pi^0|\bar s s|\pi^0 \rangle =0$ up to
high-order terms in chiral expansion.
The argument goes like this: using the first-order
perturbation theory we calculate the masses of Goldstone
bosons, for instance,$^{10}$
$$      m^2_\pi = \langle \pi |m_u \bar uu + m_d \bar dd
        + m_s \bar ss |\pi \rangle.  \eqno(33) $$
On the other hand, we can
calculate these masses from the axial current two-point
functions,
$$      m^2_\pi = -{1\over f_\pi^2}(m_u + m_d)
   \langle 0|\bar uu|0\rangle. \eqno(34) $$
Matching these two results, we have
$$ \langle \pi^0|\bar s s|\pi^0 \rangle  = 0. \eqno(35) $$
Thus, it seems difficult to find evidence that the
strange quark content is large in pion.

The $\Delta I =1/2$ rule is a solid
experimental fact. To translate this into a statement
about strange content of the nucleon and pion is
not entirely straightforward. The contributions of
penguin operators, particularly in the meson sector,
must be calculated in a more reliable way,
although they seem negligible at the scale
we consider. Our analysis in the effective theory
is made only at the tree order in chiral perturbation,
and higher order corrections could be large despite
the folklore that they are typically at the level of 30\%.
Furthermore, in the pion case we do not know the real value of
$\chi$ and the available methods for evaluating it do not
explain the $\Delta I =1/2$ rule themselves. Finally,
I should emphasize the scale dependence of strange content.
It is possible that large strange matrix elements at the scale
of 1 GeV become negligible at the scale, say, 0.2 GeV. However, besides
the untrustworthy perturbation evolution, no one knows the
scale dependence of hadron matrix elements at low energy.

I thank A. Manohar and S. Sharpe for discussions on the
penguin contributions.

\vfill
\eject
\centerline{\bf REFERENCES}
\item{1.}J. Ashman {\it et al.\/}, {\it Nucl. Phys.\/} {\bf B328} (1989) 1.
\medskip
\item{2.}the SMC collaboration at CERN, the HERMES collaboration
at HERA, the E142/E143 collaborations at SLAC, the
SAMPLE collaboration at MIT-Bates,
CEBAF proposals \#PR-91-010, \#PR-91-017, \#PR-91-004.
\medskip
\item{3.}D. B. Kaplan, {\it Phys. Lett.\/} {\bf 275} (1992)
137.
\medskip
\item{4.}J. M. Flynn and L. Randall, {\it Nucl. Phys.\/} {\bf B326} (1989) 31.
\medskip
\item{5.}H. Georgi, {\it Weak Interaction and Modern
Particle Theory} (Addison-Wesley, Reading, MA, 1984).
\medskip
\item{6.}E. Jenkins, {\it Nucl. Phys.} {\bf B375} (1992) 561.
\medskip
\item{7.}M. A. Shifman, A. I. Vainshtein, and V. J. Zakharov,
{\it Nucl. Phys.} {\bf B120} (1977) 315.
\medskip
\item{8.}R. S. Chivukula, J. M. Flynn, and H. Georgi,
{Phys. Lett.} {\bf B171} (1986) 453.
\medskip
\item{9.}S. Sharp, {\it Nucl. Phys. (Proc. Suppl.)} {\bf B20} (1990) 429.
\medskip
\item{10.}J. Gasser and H. Leutwyler, {\it Phys. Reps.\/}
{\bf 87} (1982) 77.
\vfill
\eject
\end